\documentclass[baaa]{baaa}

\usepackage[pdftex]{hyperref}
\usepackage{subfigure}
\usepackage{natbib}
\usepackage{helvet,soul}
\usepackage[font=small]{caption}

\contriblanguage{1}

\contribtype{1}

\thematicarea{7}

\title{Origin of supermassive black holes: predictions for the black hole population}

\titlerunning{Origin of supermassive black holes}

\author{
M. Liempi\inst{1},
L. Almonacid\inst{1},
D.R.G. Schleicher\inst{1}
\&
A. Escala\inst{2}
}

\authorrunning{Liempi et al.}

\contact{mliempi2018@udec.cl}

\institute{
Departamento de Astronomía, Universidad de Concepción, Chile
\and
Departamento de Astronomía, Universidad de Chile, Chile}

\resumen{La presencia de agujeros negros super masivos con corrimiento al rojo $z > 6$ plantea algunas preguntas sobre su formación y crecimiento en el universo primitivo. Debido a la construcción de nuevos telescopios como el ELT, el cual estará capacitado para observar y detectar agujeros negros super masivos, será útil derivar estimaciones teóricas de la población y comparar observaciones y predicciones de modelos en el futuro. En consecuencia nuestro principal objetivo es estimar la población de estos objetos utilizando un código semi-analítico conocido como {\sc Galacticus}, el cual es un código para la formación y evolución de galaxias donde podemos simular diferentes escenarios para la formación de agujeros negros indicando la masa inicial de estos y las condiciones de formación y evolución de los distintos componentes de las galaxias. Nuestros resultados preliminares muestran que el mecanismo principal del crecimiento de los agujeros negros es la fusión de galaxias y la acumulación de material. Además, calculamos los radios de influencia de los agujeros negros para comparar con los datos observacionales en la literatura.}

\abstract{The presence of supermassive black holes at redshift $z > 6$ raises some questions about their formation and growth in the early universe. Due to the construction of new telescopes like the ELT to observe and detect SMBHs, it will be useful to derive theoretical estimates for the population and to compare observations and model predictions in the future. In consequence our main goal is to estimate the population of SMBHs using a semi-analytic code known as {\sc Galacticus} which is a code for the formation and evolution of galaxies where we are about to include different scenarios for SMBHs formation indicating the initial mass of the black hole seed, its formation conditions and recipes for the evolution of the components of the galaxies. We found that the principal mechanism of growing SMBHs is is via galaxy mergers and accretion of matter. For the comparison of our results with observations, we calculate the radius of influence of the black hole to estimate which part of the population could be detected, leading to relations similar to the observed ones.}


\begin{document}

\maketitle

\section{Introduction}\label{S_intro}

It is known that the origin of supermassive black holes (SMBHs), whose presence in the center of most galaxies has been confirmed in the last decades \citep{kormendy2013}, is still unclear \citep{volonteri2010formation}. Moreover, the detection of nuclear star clusters (NSCs) (\cite{boker2001hst}; \cite{cote2006}) in the center of some galaxies requires a mechanism to explain the (co)-existence of SMBHs and/or NSCs in the galactic nuclei.

The best example is our Galaxy, which hosts a SMBH called Sagittarius A*. That was detected with the EHT telescope \citep{chael2023} and a NSC detected with studies of stellar orbits around the SMBH (\cite{ghez2008}; \cite{genzel2010}); \cite{gillessen2017}).

The detection of SMBHs at high redshifts (e.g., \cite{banados2018}; \cite{onoue2019}; \cite{tripodi2022}) further complicates the escenario since they must have had a short period of time to reach masses often exceeding $>10^9  ~ \mathrm{M}_\odot$ \citep{paliya2019}.

In order to solve the question regarding the origin of SMBHs, some pathways were proposed to explain how they were formed in the early universe.

The first situation proposes the existence of SMBH seeds in the early universe. A seed is a black hole with an initial mass in the range from a few hundreds to about a million solar masses \citep{volonteri2021}. The channels currently linked to the origin of black hole seeds are based on the ideas of \cite{rees1978}, starting from a “gas cloud” and depending on the physical processes that the cloud goes through (e.g. star formation, contraction, collapse and/or accretion) it could form a black hole seed.  More recent works use those ideas for the direct collapse of a primordial cloud onto a SMBH (e.g., \cite{bromm2003}; \cite{schleicher2013}).

Another channel linked to SMBH formation is driven by runaway stellar collisions in galactic nuclei. The latest results show that it is possible to form SMBHs via stellar collisions in NSCs. If a central massive object more massive than $10^8 ~ \mathrm{M}_\odot$, with a relaxation time longer than its collision time, is too dense it will collapse into a SMBH (\cite{escala2021}; \cite{vergara2022}).

In the first instance, we will focus our investigation on the evolution of SMBHs seeds scenario in {\sc Galacticus}.
\section{Methodology}
\subsection{The code}
\begin{figure}[!t]
\centering
\includegraphics[width=\columnwidth]{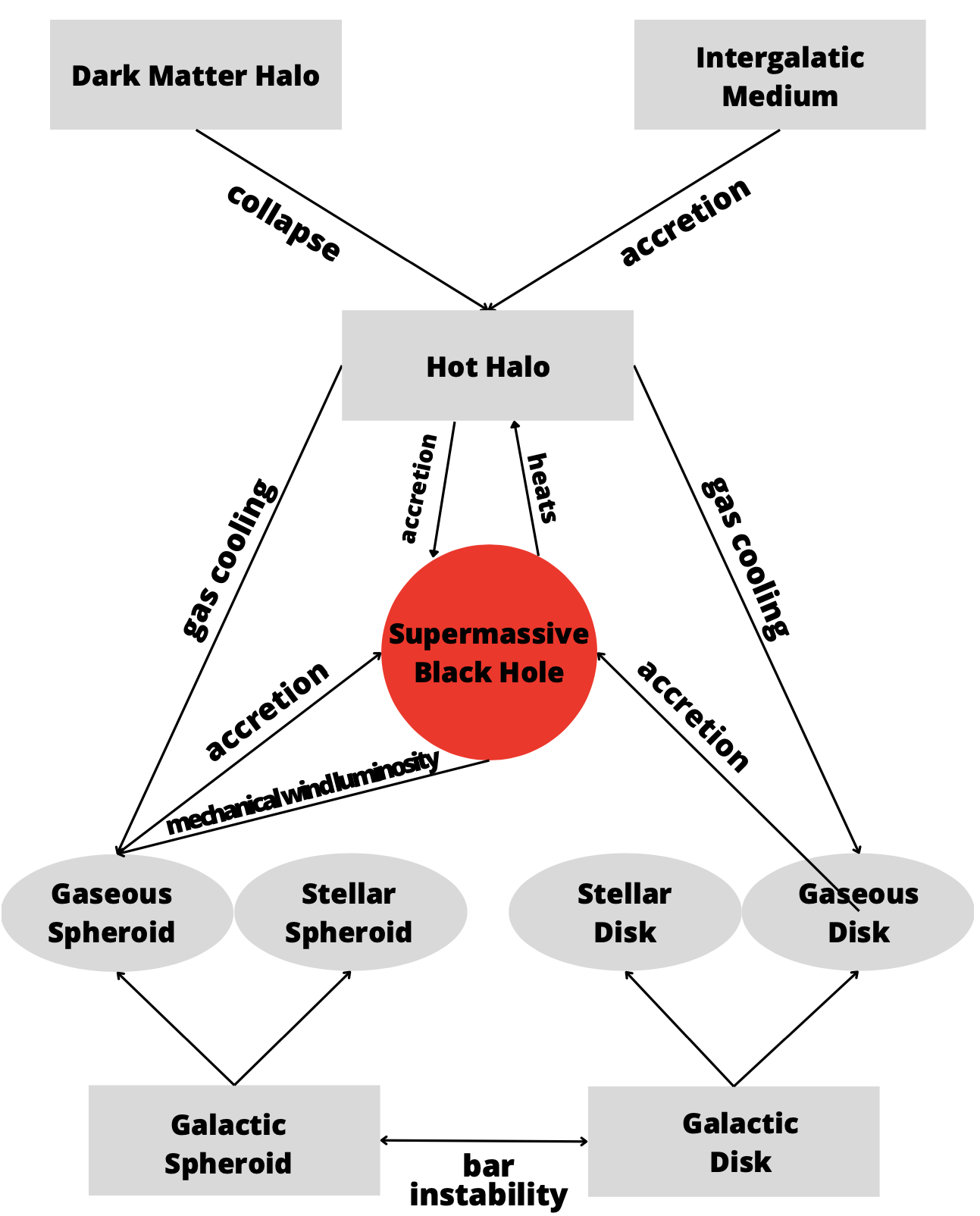}
\caption{Key interactions between the components in {\sc Galacticus}.\textit{ Not all the components exists in each galaxy}. The complete details of the interactions and the model used can be constructed seeing the manual indicated in subsection \ref{subsection2.2}.}
\label{Figura1}
\end{figure}

{\sc Galacticus} is an open source semi-analytic model to study galaxy formation and evolution written by \cite{benson2012}. It is designed to be easily expanded and explored. The advantages of {\sc Galacticus} are its computationally inexpensive cost compared to cosmological codes, allowing the fast exploration of the parameter space. The basic structure in {\sc Galacticus} is a merger tree, which consists of a linked tree of nodes. Each node represents a single dark matter halo or sub-halo (and its content) which have various properties (e.g. black hole mass, stellar mass, gas mass, etc.). For more details about the components and the interaction between each other see Fig. \ref{Figura1}.
\subsection{Merger tree sample building}\label{subsection2.2}
We use the merger tree task of Galacticus to build our merger trees where we start from a uniformly distributed sample at $\mathrm{z} = 0$ with masses between $10^{6} -10^{15}~ \mathrm{M}_\odot$. The initial distribution is specified at $z=0$ for the reason that {\sc Galacticus} works reconstructing the history of merger trees following it backward in time. The internal merger tree builder uses the algorithm of \cite{cole2000} with a minor modification and requires some description of the branching probabilities in trees. We choose the branching probability distribution  of \cite{parkinson2008}  and a sampling rate for the mass evolution of the halo given by a power-law with the exponent $\alpha=1$. All the details about the models available can be found on {\sc GITHUB}\footnote{\url{https://github.com/galacticusorg/galacticus}}.
\subsection{Supermassive black holes}
 In this section we describe briefly the physical meaning of the parameter set in model A \& B. The initial conditions for the SMBH initialization are given in Table \ref{tabla1}. The evolution of the SMBH is given by:
\begin{equation}
M_\bullet = (1-\epsilon_\mathrm{radiation}-\epsilon_\mathrm{jet})\dot{M}_0,
\end{equation}
where $\epsilon_\mathrm{radiation}$ is the radiative efficiency of the accretion flow feeding the black hole, $\epsilon_\mathrm{jet}$ is the efficiency with which accretion power is converted into jet power and $\dot{M}_0$ is the rest mass accretion rate.{\sc Galacticus} assumes the rest mass accretion as a Bondi-Hoyle-Lyttlelon accretion rate from the spheroid gas reservoir.
The black hole is assumed to cause feedback in two ways (see Fig. \ref{Figura1}):
\begin{itemize}
\item The first way is a radio mode, where any jet power from the black hole-accretion disk system is included in the hot halo heating rate.
\item The other way is given by the addition of the mechanical wind luminosity of \cite{ostriker2010} to the gas component of the spheroid. The expression used is
\end{itemize}
\begin{equation}
L_\mathrm{wind}=\epsilon_\mathrm{wind}\epsilon_\mathrm{radiation}\dot{M}_0c^2,
\end{equation}
with $\epsilon_\mathrm{wind}$ the black hole wind efficiency. The initial value of  $\epsilon_\mathrm{wind}$ is fixed to be equal to the best fit model value described in {\sc Galacticus GITHUB} website.

\section{Preliminary results}
We calculate the radius of the sphere of influence of the SMBHs using the following definition
\begin{equation}
r_\mathrm{influence} = \frac{GM_\bullet}{\sigma^2},
\end{equation}
where $G$ is the gravitational constant, $M_\bullet$ is the mass of the black hole and $\sigma$ stellar velocity dispersion of the host bulge following the  definition of \cite{kormendy2013}.

\begin{figure}[!t]
\centering
\includegraphics[width=\columnwidth]{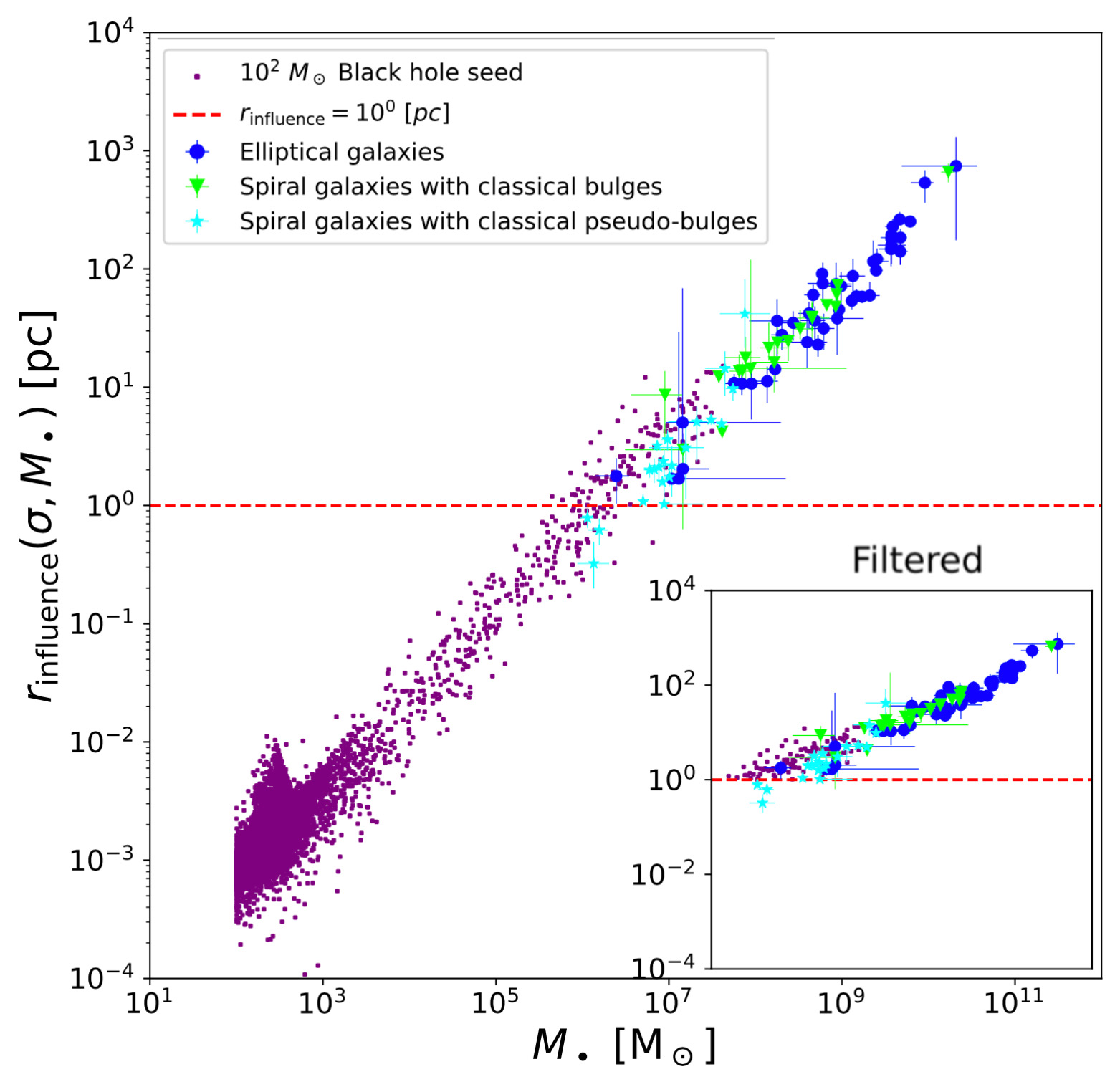}
\caption{Figure 3: Comparison between the influence radius for simulated data and observed data. We also show how the population changes when we select SMBHs with $r_\mathrm{influence} > 1~ \mathrm{pc}$. The observational data was extracted from \cite{kormendy2013}, where classical bulges are defined as an elliptical galaxy, except that they are embedded in disks, while pseudo-bulges are classical bulges which have properties that are more disk-like than those of classical bulges.}
\label{Figura3}
\end{figure}

We define a critical value $r_\mathrm{influence} = 1 ~ \mathrm{pc}$, calculated with the data provided in \cite{kormendy2013}, in order to discriminate SMBHs which could be detected at z=0 via stellar/gas dynamics. These results are summarized in Fig. \ref{Figura3} where it is possible to see, how a better resolution can impact in the detection of SMBHs with $r_\mathrm{influence} < 1 ~\mathrm{pc}$.

\begin{figure}[!t]
\centering
\includegraphics[width=\columnwidth]{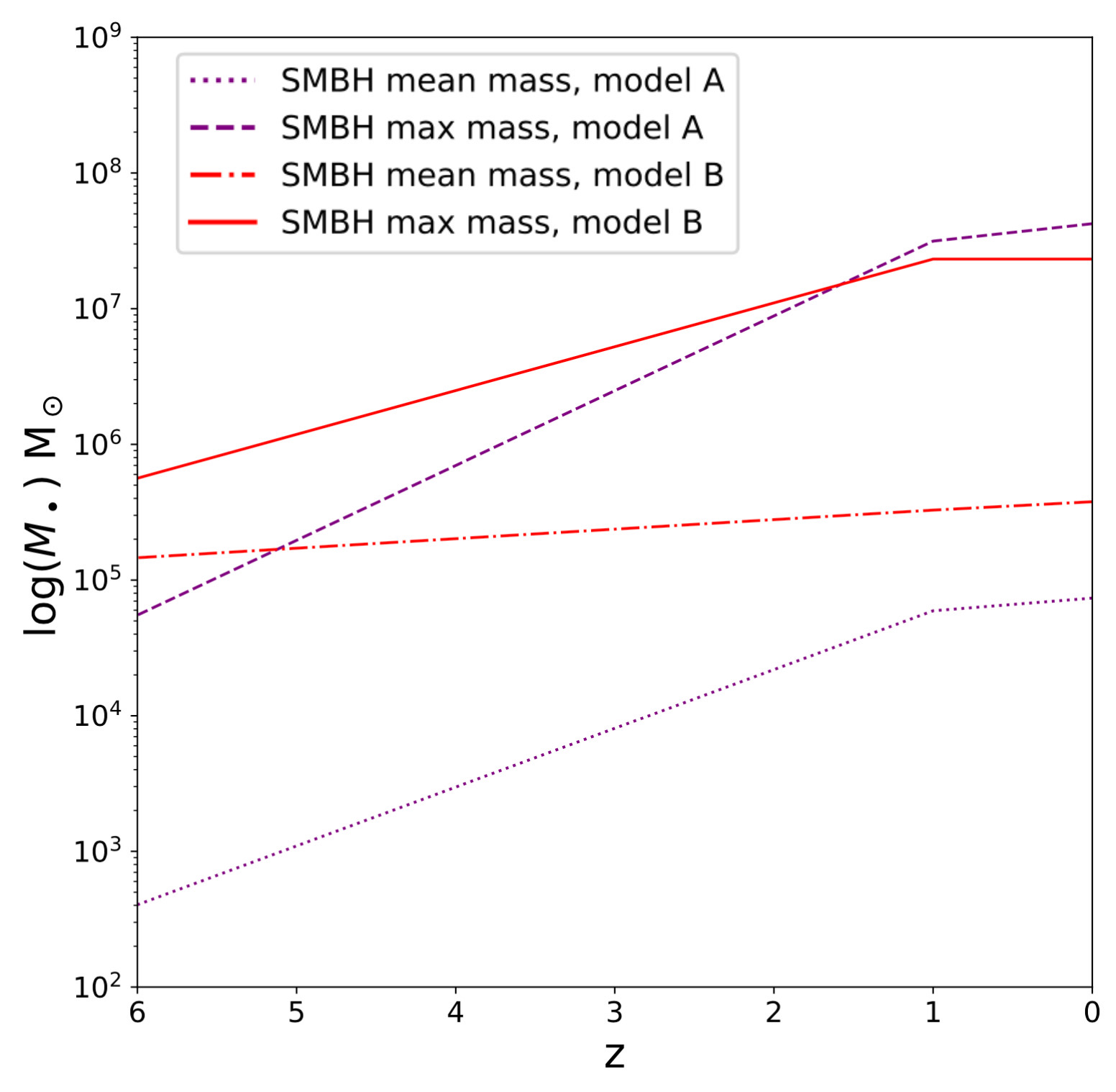}
\caption{Mass evolution of models A \& B in function of redshift. We show the mean mass and the maximum mass of the SMBHs at a given z. Model A represents a light seed while model B represents a heavy seed.}
\label{Figura2}
\end{figure}

We also explore the mass evolution for the model A and B (see Fig. \ref{Figura2}), where we found a similar evolution independent of the  initial black hole seed mass.

\begin{table}[!t]
\centering
\caption{Relevant parameters used to initialize the black hole component in {\sc Galacticus}.}
\begin{tabular}{lccc}
\hline\hline\noalign{\smallskip}
\!\!Model & \!\!\!\! $\epsilon_\mathrm{wind}$ & \!\!\!\!Seed mass$~\mathrm{M}_\odot$& \!\!\!\!Mass resolution$~\mathrm{M}_\odot$\!\!\!\!\\
\hline\noalign{\smallskip}
\!\!A  &  0.0024 & $10^2$ & $10^{10}$\\
\!\!B  & 0.0024  & $10^5$ & $10^{10}$\\
\hline
\end{tabular}
\label{tabla1}
\end{table}

\section{ Conclusion and future work}
Although our work is preliminarily in agreement with \cite{volonteri2021}, who concludes that accretion is the first source for the SMBHs growth, and mergers play a secondary role in the mass budge of SMBHs, we need to explore the dependence on the mass resolution to improve our data analysis and to made the statistical work for the SMBH population.
We found that observations detected SMBHs with $r_\mathrm{influence} \gtrsim 1~\mathrm{pc}$. It is evident that the number of the SMBHs detections could be higher  if the resolution to detect dynamically the local SMBHs increases, and of course, this is something that we will explore in more detail to predict the SMBH population.
Finally, in the future we plan to implement a black hole formation scenario based on collisions in NSCs.

\begin{acknowledgement}
 We gratefully acknowledge support by the ANID BASAL projects ACE210002 and FB210003, as well as via the Millenium Nucleus NCN19-058 (TITANs).
\end{acknowledgement}


\bibliographystyle{baaa}
\small
\bibliography{bibliografia}
 
\end{document}